\documentclass[conference]{IEEEtran}
\IEEEoverridecommandlockouts

\usepackage{cite}
\usepackage{amsmath,amssymb}
\usepackage{booktabs}
\usepackage{multirow}
\usepackage{url}
\usepackage{xcolor}
 \usepackage{graphicx} 
\begin{document}

\title{The tttAI System for the TSA-ASR Task of the SmartGlasses Challenge 2026}

\author{
\IEEEauthorblockN{Xuanji He$^{1}$, Gaoyang Dong$^{1}$, Xiaoxiao Li$^{1}$, Minchuan Chen$^{1}$, Fengjie Zhu$^{1}$}
\IEEEauthorblockA{$^{1}$Shenzhen Transsion Holdings Co., Ltd, China\\
\{xuanji.he, gaoyang.dong, xiaoxiao.li7, minchuan.chen, fengjie.zhu\}@transsion.com}
}

% \author{
% \IEEEauthorblockN{Xuanji He}
% \IEEEauthorblockA{
% Shenzhen Transsion Holdings Co., Ltd, China\\
% Email: hexuanji99@gmail.com}
% \and
% \IEEEauthorblockN{Gaoyang Dong}
% \IEEEauthorblockA{
% Shenzhen Transsion Holdings Co., Ltd, China\\
% Email: gaoyang.dong@transsion.com}

% \and
% \IEEEauthorblockN{Xiaoxiao Li}
% \IEEEauthorblockA{
% Shenzhen Transsion Holdings Co., Ltd, China\\
% Email: xiaoxiao.li7@transsion.com}
% }
\maketitle

\begin{abstract}
This paper presents the tttAI system submitted to the TSA-ASR task of the SmartGlasses Challenge 2026, evaluated on both two-person dialogues (Track~1) and multi-party meetings (Track~2).
The task requires time-stamped speaker-attributed speech recognition from smart-glasses recordings. 
This is particularly challenging due to long-form audio, multiple speakers, and frequent overlapping speech.
We proposed a cascaded architecture consisting of speaker diarization, overlap detection, target-speaker extraction, post-processing, and automatic speech recognition. The diarization module extracts features via WavLM‑Large, performs frame‑wise speaker classification with a Conformer encoder, and then generates global speaker segments through embedding clustering. For overlapped regions, we apply a WeSep-based target-speaker extraction model with ECAPA-TDNN speaker embeddings. 
When the extraction is unreliable, a dominant-speaker fallback strategy is used.
The final system uses FireRedASR2-AED with the first microphone channel. 
The submitted system has a total parameter count of approximately 1.53B.
On Track~1, our system achieves a tcpCER of 7.10\%. On Track~2, it achieves a tcpCER of 34.04\% and ranks second on the leaderboard.
\end{abstract}


\section{Introduction}
% Smart glasses are rapidly emerging as the next-generation wearable computing platform.
% They promise always-on, eyes-free interaction that seamlessly blends digital information with the physical world. 
% Unlike smartphones or smart speakers, smart glasses operate in highly dynamic, multi-talker environments.
% These range from dyadic face-to-face conversations to multi-participant meetings. 
% The device must understand not just what is being said, but also who is speaking. 
% Robust speech recognition in the presence of interfering speakers is therefore a fundamental requirement for any practical smart glasses system ~\cite{smartglasses2026}.

% Conventional automatic speech recognition (ASR) systems are typically designed for single-speaker, close-talk scenarios. 
% Their performance severely degrades when multiple speakers overlap or alternate in the same acoustic scene.
% Time-stamped speaker-attributed ASR (TSA-ASR) directly addresses this limitation. 
% Given a short enrollment utterance or other speaker-discriminative cues from the target user, the system transcribes only the target speaker's speech and ignores all competing voices.
% TSA-ASR thus stands as one of the core enabling technologies for personalized, hands-free interaction on smart glasses.

Smart glasses are emerging as a next-generation wearable platform for always-on, eyes-free interaction.
They often operate in dynamic multi-talker scenarios, such as conversations and meetings, where the system must recognize not only what is said but also who is speaking~\cite{smartglasses2026}.
Conventional ASR systems degrade severely under overlapping speech.
Time-stamped speaker-attributed ASR (TSA-ASR) addresses this challenge by transcribing timestamped speech from the target speaker while ignoring competing voices, making it essential for practical smart-glasses interaction.

The SmartGlasses Challenge 2026 benchmarks TSA-ASR in realistic wearable scenarios, offering two sub-tracks: two-person dialogues (Track~1) and multi-speaker meetings (Track~2)~\cite{smartglasses2026}. 
% Audio is captured by a smart glasses microphone array, and the task is to transcribe a target speaker given a short enrollment clip.
Audio is captured by a smart glasses microphone array, and the task is to transcribe a target speaker by leveraging speaker-specific cues extracted from the recording itself.
Dialogues involve tight turn-taking and frequent overlaps, while meetings add more speakers, reverberation, and longer sessions.
We address both tracks, with primary evaluation on Track~2.
The main metric is tcpCER (time-constrained minimum permutation WER, 5-second collar), with cpWER and DER as diagnostics.

We proposed a cascaded architecture rather than a single end-to-end model. 
A neural speaker diarization module (DiariZen-style EEND \cite{han2024diarizen} with WavLM-Large \cite{chen2022wavlm} and a Conformer backend \cite{gulati2020conformer}) estimates speaker turns and overlap regions. 
For overlapped speech, a WeSep-based TSE model \cite{wang2024wesep} with a band-split RNN separator \cite{luo2023bsrnn} and ECAPA-TDNN speaker conditioning \cite{desplanques2020ecapa} extracts target speaker signals using enrollment segments selected from nearby non-overlapped speech. 
A dominant-speaker fallback is applied when the extracted signal is too similar to the mixture: the dominant speaker retains the mixture segment while non-dominant speakers drop the overlap. 
The resulting speaker-conditioned segments are decoded by FireRedASR2-AED and output in STM format. 
This modular design lets us independently address diarization-induced speaker confusion, insertion errors from failed separation, and ASR robustness on wearable recordings.

\begin{figure*}[htbp]
    \centering
    \includegraphics[width=\linewidth]{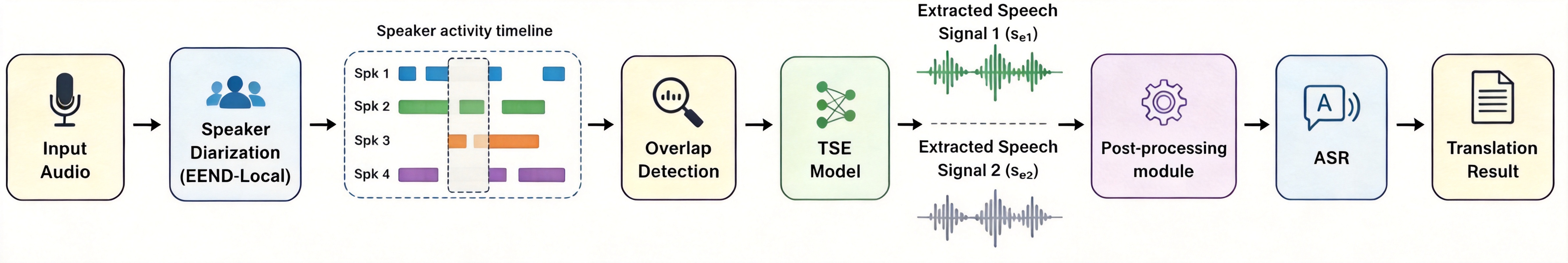}
    \caption{Overview of the proposed system framework. Given an input audio recording, the system first performs speaker diarization using a local EEND-style model to produce speaker activity timelines. Overlapped speech regions are then detected from the diarization output and processed by the target-speaker extraction (TSE) model. The extracted signal is further examined by a post-processing module to handle unreliable extraction cases. Finally, the post-processed speaker-specific segments are passed to the ASR backend to generate the translation result.}
    \label{fig:system_framework}
\end{figure*}

\section{Data and Task}

The SmartGlasses Challenge 2026 TSA-ASR task requires transcribing the target speaker's speech with utterance-level timestamps from multi-channel smart-glasses recordings. 
The dataset, captured by a four-channel microphone array, is divided into Track~1 (short dyadic dialogues) and Track~2 (long multi-party meetings). 
Track~2 has longer sessions, more speakers (3--8, avg.\ 4.3), and a higher overlap ratio (13.37\% vs.\ 7.59\% in Track~1), making it substantially more challenging.
The speech is predominantly spontaneous Mandarin Chinese with occasional English code-switching, effectively monolingual.

Table~\ref{tab:dataset} summarizes the training and development set statistics for both tracks.
We use only the first microphone channel, as it yields the best development tcpCER among the four single-channel candidates. 
For local tuning, we merge the official development partitions within each track into a single validation set, resulting in 81 and 36 development recordings for Track~1 and Track~2, respectively. 
All reported results are evaluated on these merged development sets. Outputs are converted to STM format and evaluated with the official tcpCER protocol based on MeetEval~\cite{MeetEval23}.

\begin{table}[htbp]
\centering
\caption{Dataset statistics.}
\label{tab:dataset}
\begin{tabular}{lcccc}
\toprule
Track & Split & Utts. & Dur. (h) & Avg. Dur. (min) \\
\midrule
\multirow{2}{*}{Track1} & train & 387 & 33.92 & 5.26  \\
                         & dev   & 81  & 6.86  & 5.08  \\
\multirow{2}{*}{Track2} & train & 130 & 41.83 & 19.30 \\
                         & dev   & 36  & 11.51 & 19.19 \\
\bottomrule
\end{tabular}
\end{table}

\section{System Description}

Fig~\ref{fig:system_framework} illustrates the overall system framework. 
Given an input audio recording, the system first performs speaker diarization to obtain speaker-specific activity timelines. 
Overlapped speech regions are then detected from 
the diarization output and processed by a target-speaker extraction (TSE) model. 
A post-processing module is further applied to handle unreliable extraction results 
before the resulting speaker-specific segments are sent to the ASR backend.

\subsection{Speaker Diarization}

The speaker diarization module follows a local EEND-style design 
\cite{fujita2019eend,horiguchi2021eendeda}. The input audio is processed in 16-second chunks. WavLM-Large is used to extract self-supervised frame-level representations, which are then projected to a 256-dimensional space and passed to a four-layer Conformer encoder. The model predicts frame-level speaker activity labels, with a maximum of four speakers in each chunk. Global speaker labels are 
obtained by clustering speaker representations, followed by diarization post-processing.

The training data include AMI 
\cite{carletta2005ami}, AliMeeting \cite{yu2022m2met}, AISHELL-4 
\cite{fu2021aishell4}, NOTSOFAR-1 \cite{vinnikov2024notsofar}, MagicData-RAMC 
\cite{yang2022ramc}, VoxConverse \cite{chung2020voxconverse}, DIHARD 
\cite{ryant2020dihard}, MLC-SLM \cite{mu2026summary}, and the official SmartGlasses data. During 
training, WavLM is frozen for the first two epochs, after which WavLM and the 
diarization backend are jointly optimized with separate learning rates.

% During inference, diarization segments shorter than 0.2 seconds are removed. 
% Adjacent segments from the same speaker are then merged, with a maximum segment 
% duration of 120 seconds. This post-processing step keeps the ASR inputs sufficiently 
% long while avoiding excessively long segments.

\subsection{Overlap Detection}

Based on the speaker activity timelines produced by the diarization module, we identify overlapped regions where multiple speakers are active (at most two in our setting). These regions are used to determine whether additional target-speaker extraction is required.
Non-overlapped regions are directly assigned to the corresponding speaker. Whereas overlapped regions are forwarded to the TSE module for speaker-specific speech extraction.
This overlap detection step is important because directly sending overlapped speech to ASR for each active speaker can lead to duplicated hypotheses and increased insertion errors. 
Therefore, only the detected overlapped regions are processed by the extraction module, while the original non-overlapped speech segments are 
preserved.

\subsection{TSE Model}

For each detected overlap involving two speakers, we apply target-speaker 
extraction to obtain speaker-specific speech signals. For every active speaker in 
an overlapped region, a speaker enrollment signal is selected from the nearest 
non-overlapped segment belonging to the same speaker. The enrollment segment is 
required to be at least 5 seconds long and is capped at 20 seconds.

The TSE model is implemented using WeSep \cite{wang2024wesep}. We used a 
band-split RNN separator \cite{luo2023bsrnn} together with ECAPA-TDNN speaker 
embeddings \cite{desplanques2020ecapa}.
For each overlapped region, we extract two speaker-specific signals, denoted as $s_{e1}$ and $s_{e2}$.

% \subsection{Post-processing Module}

% Although the TSE model can separate speaker-specific signals from overlapped 
% regions, extraction errors may still occur, especially when the enrollment segment 
% is noisy or when the target speaker is weak in the mixture.
% To improve robustness, 
% we introduce a failure-aware dominant-speaker post-processing strategy.

% Let $s_e$ denote the extracted overlap signal and $s_m$ denote the original mixture. 
% If the cosine similarity between the extracted signal representation and the 
% mixture representation is above a predefined threshold, the extraction is considered 
% unreliable. This indicates that the extracted signal may still be too similar to the 
% original mixture and may not provide a clean target-speaker signal.

% In such cases, we compare the extracted signal with the enrollment embeddings of 
% all overlapped speakers. The speaker with the highest similarity is treated as the 
% dominant speaker. The dominant speaker keeps the original mixture segment in the 
% corresponding ASR input, while non-dominant speakers remove that overlapped region 
% from their ASR segments. This strategy reduces duplicated recognition results and 
% helps suppress insertion errors caused by unreliable extraction. The similarity 
% threshold is tuned on the development set and is set to 0.55.

\subsection{Post-processing Module}

Although the TSE model can separate speaker-specific signals from overlapped regions, extraction errors may still occur. To improve robustness, we introduce a failure-aware dominant-speaker post-processing strategy.
For each overlapped region, the TSE model extracts speaker-specific signals for all active speakers (at most two in our setting).
% Let $s_{e1}$ and $s_{e2}$ denote the extracted signals for the two speakers, respectively. 
We compute the cosine similarity between $s_{e1}$ and $s_{e2}$ embedding representations, extracted from the same speaker embedding model used for enrollment. 
If the similarity exceeds a predefined threshold, the extraction is considered unreliable, indicating that the separator failed to adequately disentangle the two speakers.

In such unreliable cases, we fall back to a dominant-speaker assignment strategy. 
Instead of using the poorly separated extracted signals, we compare the original mixture against the enrollment embeddings of each overlapped speaker.
The speaker whose enrollment embedding yields the highest similarity to the mixture is designated as the dominant speaker. 
The dominant speaker retains the original mixture segment in the corresponding ASR input, while the other speaker removes that overlapped region from their ASR segments.
% The similarity threshold is tuned on the development set and is set to 0.55.

\begin{table}[t]
\centering
\caption{Diarization system comparison on Track 2 development set.}
\label{tab:diarization}
\small
\begin{tabular}{lrrrr}
\toprule
System & MISS & SPK & DER & tcpCER \\
\midrule
Open-source DiariZen & 8.79 & 6.69 & 22.44 & 27.26 \\
SmartGlasses-only DiariZen & \textbf{5.39} & 8.43 & 21.07 & 29.74 \\
Open-data + SmartGlasses & 7.21 & \textbf{4.57} & \textbf{18.62} & \textbf{22.88} \\
\bottomrule
\end{tabular}
\end{table}

\begin{table}[t]
\centering
\caption{TSE model comparison on Track 2 development set.}
\label{tab:tse}
\small
\begin{tabular}{lrrr}
\toprule
TSE model  & tcpCER & WER \\
\midrule
Large-scale SV-trained TSE  & 24.69 & 20.60 \\
SmartGlasses-adapted TSE  & \textbf{23.64} & \textbf{19.80 }\\
\bottomrule
\end{tabular}
\end{table}

\begin{table}[t]
\centering
\caption{Overlap strategy comparison on Track 2 development set, based on experimental verification with Qwen3-ASR-1.7b.}
\label{tab:scheme}
\small
\begin{tabular}{lrrrr}
\toprule
Strategy & MISS & SPK & DER & tcpCER \\
\midrule
Native segmentation & 5.07 & 5.16 & 17.53 & 26.64 \\
Drop overlap & 27.33 & \textbf{2.06} & 34.54 & 32.88 \\
Assign overlap & 15.16 & 3.56 & 23.95 & 25.78 \\
TSE overlap & \textbf{5.07} & 5.16 & \textbf{17.53} & 24.69 \\
TSE dominant overlap & 7.21 & 4.57 & 18.62 & \textbf{22.88} \\
\bottomrule
\end{tabular}
\end{table}

\begin{table}[t]
\centering
\caption{Channel and ASR comparisons on Track 2 development set. The channel results are obtained using FireRedASR2-AED under the TSE overlap configuration. DER collar is 0s}
\label{tab:channel_asr}
\small
\begin{tabular}{llrr}
\toprule
Factor & System & DER & tcpCER \\
\midrule
\multirow{4}{*}{Channel}
& ch1 & \textbf{17.53} & \textbf{23.61} \\
& ch2 & 17.81 & 24.09 \\
& ch3 & 18.12 & 24.32 \\
& ch4 & 18.50 & 25.25 \\
\midrule
\multirow{1}{*}{Baseline}
&Vibevoice-ASR &  32.50 & 33.77 \\
\midrule
\multirow{2}{*}{ASR}
& Qwen3-ASR-1.7B & - & 22.88 \\
& FireRedASR2-AED & - & \textbf{21.73} \\
\bottomrule
\end{tabular}
\end{table}

\subsection{ASR Model Selection}

The post-processed speaker-specific segments are finally passed to the ASR backend. 
We evaluate Qwen3-ASR-1.7b \cite{shi2026qwen3} and FireRedASR2-AED \cite{xu2026fireredasr2s}. Qwen3-ASR-1.7b is used for fast development experiments due to its inference speed, which enables rapid iteration. For the final submission, we adopt FireRedASR2-AED, as it achieves lower tcpCER.

\section{Experiments}

\subsection{Model parameter sizes}
 We explicitly declare that the total parameter size does not exceed 35B. The diarization model, consisting of WavLM-Large and a Conformer backend, contains 321.8M parameters. The TSE model, based on BSRNN with ECAPA conditioning, contains 27.6M parameters. The FireRedASR2-AED, contains 1,183M parameters. Therefore, the total number of parameters is 1,532.4M. 
 
\subsection{Experimental Setup}
% The diarization model is trained for 30 epochs. WavLM is frozen for the first two epochs and then jointly optimized with the diarization backend. The learning rates are set to 1e-5 for WavLM and 1e-3 for the remaining components. Noise augmentation is introduced progressively, with single-noise augmentation from epoch 6 and mixed-noise augmentation from epoch 15. The TSE model is first trained on a speaker-verification corpus with a learning rate of 1e-3, and then fine-tuned on the official SmartGlasses train data with a learning rate of 1e-4. The cosine-similarity threshold for detecting unreliable extraction is tuned on the development set and set to 0.55.
The diarization model is trained for 30 epochs. WavLM is frozen for the first two epochs and then jointly optimized with the backend, using learning rates of 1e-5 and 1e-3 for WavLM and the remaining components, respectively. Noise augmentation starts from epoch 6 for single-noise and epoch 15 for mixed-noise conditions. The TSE model is pre-trained on a speaker-verification corpus and fine-tuned on the official SmartGlasses training data with learning rates of 1e-3 and 1e-4, respectively. The unreliable-extraction threshold is tuned on the development set and set to 0.55.
 
\subsection{Development Results}
Table \ref{tab:diarization} compares diarization systems.
The open-source baseline \cite{han2025fine} has the highest DER and tcpCER.
The SmartGlasses-only model reduces MISS but has high speaker confusion and poor tcpCER. The final open-data plus SmartGlasses model has the best overall tcpCER among the three systems.
Table \ref{tab:tse} compares the two TSE models using the same diarization and ASR setup. The SmartGlasses-adapted TSE model has almost unchanged DER but reduces tcpCER from 24.69\% to 23.64\%, showing that adaptation mainly improves the ASR quality of separated overlap speech.

Table~\ref{tab:scheme} compares overlap handling strategies. Native segmentation directly feeds diarization outputs to ASR. Dropping all overlap severely increases MISS. Assigning overlap to a single speaker reduces insertions but loses much target speech. For TSE overlap, we extract speaker-specific signals for each speaker in overlap and feed them to ASR directly. Our proposed TSE dominant overlap adds a fallback: when the two extracted signals are too similar, we assign the original mixture to the speaker whose enrollment embedding is most similar to it, and drop that region for the other speaker. 
This achieves the best development-set result, reducing tcpCER from 24.69\% to 22.88\%.
Table \ref{tab:channel_asr} summarizes channel and ASR comparisons. Among the four single channels, channel 1 gives the lowest tcpCER. With the same diarization and TSE configuration, FireRedASR2-AED reduces tcpCER from 22.88\% to 21.73\% compared with Qwen3-ASR-1.7b, so it is used in our final system.

\subsection{Official Test Result}
% The final Track 2 Task 1 submission uses the open-data plus SmartGlasses DiariZen model, SmartGlasses-adapted TSE, dominant-speaker overlap handling, FireRed ASR, and the first microphone channel. On the official test set, the system obtains 34.04 tcpCER and ranks second on the Track 2 Task 1 leaderboard.
On the official test set, the final system achieves a tcpCER of 7.10\% on Track~1, ranking fifth on the Track~1 leaderboard, and a tcpCER of 34.04\% on Track~2, ranking second on the Track~2 leaderboard. 
% These results demonstrate the effectiveness of the proposed cascaded TSA-ASR system across both two-person dialogue and multi-party meeting scenarios.

\section{Conclusion}
% We presented the tttAI system for Track 2 Task 1 of the SmartGlasses Challenge 2026. The system uses a cascaded speaker-attributed ASR pipeline with neural diarization, target-speaker extraction, failure-aware overlap handling, and FireRed ASR. Development experiments show that the most important improvements come from robust diarization trained with open data and SmartGlasses data, SmartGlasses-adapted TSE, and dominant-speaker fallback for failed overlap extraction. The final system achieves 34.04 tcpCER on the official test set and ranks second.

% This paper presented the tttAI system for the TSA-ASR task of the SmartGlasses Challenge 2026, covering both two-person dialogues and multi-party meetings.
% The system follows a cascaded speaker-attributed ASR pipeline with diarization, overlap-aware target-speaker extraction, post-processing, and FireRed ASR.
% The system uses diarization trained on both public and SmartGlasses data. 
% This is combined with a SmartGlasses-adapted TSE model and a dominant-speaker fallback strategy for unreliable extraction.
% Official test results show that our system achieves a tcpCER of 7.10\% on Track~1 and a tcpCER of 34.04\% on Track~2, ranking second on the Track~2 leaderboard.

This paper presented the tttAI system for the TSA-ASR task of the SmartGlasses Challenge 2026.
The system follows a cascaded speaker-attributed ASR pipeline with diarization, overlap-aware target-speaker extraction, post-processing, and FireRedASR2-AED.
Official test results show that our system achieves a tcpCER of 7.10\% on Track~1 and a tcpCER of 34.04\% on Track~2, ranking second on the Track~2 leaderboard.

\bibliographystyle{IEEEtran}
\bibliography{smartglasses_tttai_refs}

\end{document}